\documentclass[
    ,final            % use final for the camera ready runs
  ]
  {aipproc}

\layoutstyle{6x9}
\usepackage{epsfig}
\begin{document}

\title{Deconfinement$^1$}\footnote{Summary of the Deconfinement Session of the Quark Confinement and Hadron Spectrum VII Conference. Ponta Delgada, Azores(Portugal) }

\classification{<Replace this text with PACS numbers; choose from this list:}
               
\keywords      {<Enter Keywords here>}

\author{C. Pajares}{
  address={Instituto Galego de F\'{\i}sica de Altas Energ\'{\i}as, Departamento de F\'{\i}sica de Part\'{\i}culas.
\\Universidade de Santiago de Compostela. 15782 Santiago de Compostela, Spain.}
}

\begin{abstract}

This is an attempt to summarize the talks given at the session on
Deconfinement in the Conference ``Quark Confinement and hadron spectrum''. This talk covers the following topics: Elliptic flow
and evidence of nearly perfect fluid of the created matter; High Transverse momentum production,
propagation of jets and energy loss; Heavy quarqonia in dense QCD matter; Phase transition, Multiplicity
fluctuations and long range correlations; Multiparticle production and thermalization.

\end{abstract}

\maketitle

In the last few years a large progress has been done in the knowledge of the deconfined phase
of QCD. The SPS data already displayed several facts that hinted at the onset of Quark Gluon Plasma (QGP)
formation. The RHIC data have conclusively discovered a striking set of new phenomena. Most of these data
have been extensively discussed in the different talks of the session of Deconfinement.

The space limits prevent me from describing all the reported exciting developments, so I will concentrate on some
of them.

\section {Elliptic flow}

The flow pattern of thousand of particles produced in a heavy ion reaction is the main observable
used to look for collective behaviour and its properties. These properties test the conditions
necessary for the obtention of QGP. One is the degree of thermalization. The evolution of the 
matter from the initial conditions can be computed by means of relativistic hydrodynamics
if local equilibrium is maintained. These equations can be further approximated by perfect fluid
equation when the viscosity correction can be neglected. The second condition is the validity
of the equation of state, numerically determined from QCD. The data on elliptic flow show evidence
that a fast thermalization is reached at RHIC energy, compatible with a soft equation of state
and a low viscosity. The matter created at RHIC behaves as a perfect fluid \cite{1}.

The elliptic flow, $v_2=<p{^2_x}-p{^2_y}/p^2_t>$, results from pressure gradients developed in
the initial almond-shaped collision zone. That is, the initial transverse coordinate space
anisotropy of the collision zone or eccentricity $\epsilon=<(y^2-x^2)/(y^2+x^2)>$ is converted,
via hadronic or partonic interactions into an azimuthal momentum anisotropy. Elliptic flow
self-quenches due to expansion of the collision zone, therefore in order to achieve relatively
large $v_2$ a fast thermalization is required (see Lisa and Bai-Yuting talks~ \cite{2}~ and~\cite{3}).
Therefore it is expected that the elliptic flow scaled by the eccentricity should be proportional
to the density of scatterings, $\frac{dN}{dy}\frac{1}{S}$, being $S$ the overlapping collision area.
This is well satisfied as fig. 1 shows. It is also shown the hydrodynamics result for a perfect
fluid \cite{4}, which only is reached for central Au-Au collisions. As a for LHC, for central Au-Au collisions
$\frac{1}{s}\frac{dN}{dy}\simeq 80$ it is expected a change on the shape of the curve becoming flat.
In fig. 2 we show the agreement of the observed hadron mass dependence of $v_2$ with the
hydrodynamics predictions below $p_t=1$ GeV/c. This result shows that there is a common
collective flow velocity.In fig. 3 it is shown the scaling law of $v_2/n$ versus 
$p_T/n$, $n$ being the number of quarks of the respective hadrons. 
This scaling law was predicted by coalescence models
suggesting that the collective flow is at the partonic stage. The coalescence models,
also explain naturally the differences between the inclusive cross sections for
baryons and mesons at intermediate transverse momentum as it was explained in
the Hippolyte talk \cite{5}.
The hydrodynamics perfect fluid prediction \cite{6} for the higher azimuthal momentum $v_4$ is
$v_4=\frac{1}{2}v{^2_2}$, but the data for $\pi^\pm$ and $p$ and $\bar{p}$ in minimum
bias Au-Au collisions gives a factor $3/2$ instead $1/2$.
However in a perfect fluid model \cite{7} it is obtained the scaling law $v_4=v{^2_2}/2 + k_4 y^4_T$,
which is in agreement with data. $k_4$ is a constant depending on the mass of
the particle and $y_T=\frac{1}{2}log(m_T + p_T)/(m_T-p_T)$.

 \noindent
\begin{minipage}[t]{6cm}
\centering\epsfig{figure=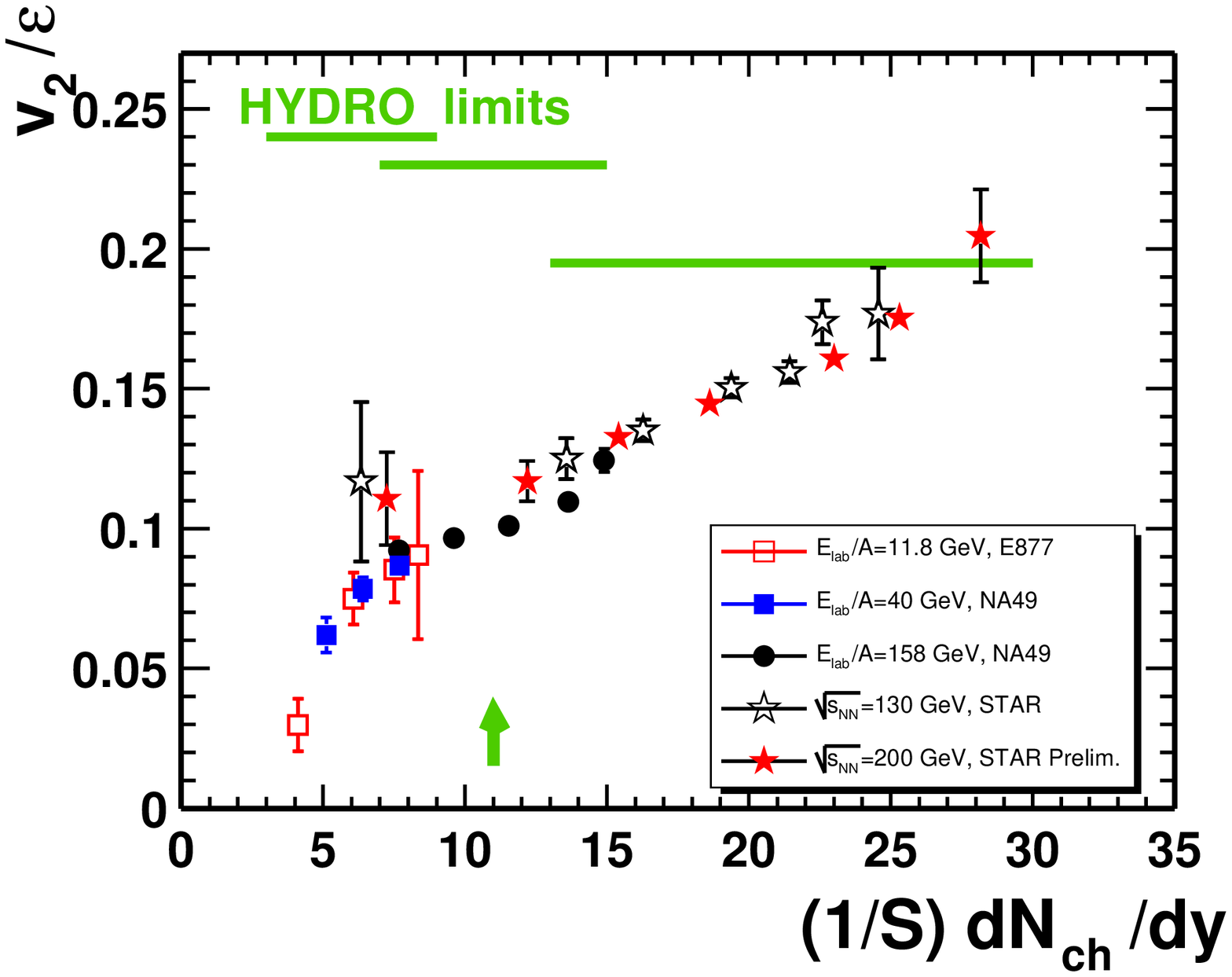,width=6cm}
{\bf FIGURE 1.} \footnotesize{$\frac{\nu}{\epsilon}$ versus
  $\frac{1}{S}\frac{dN_{ch}}{dy}$ for different energies and centralities.}
\end{minipage} 
\hfill
\begin{minipage}[t]{6cm}
\centering\epsfig{figure=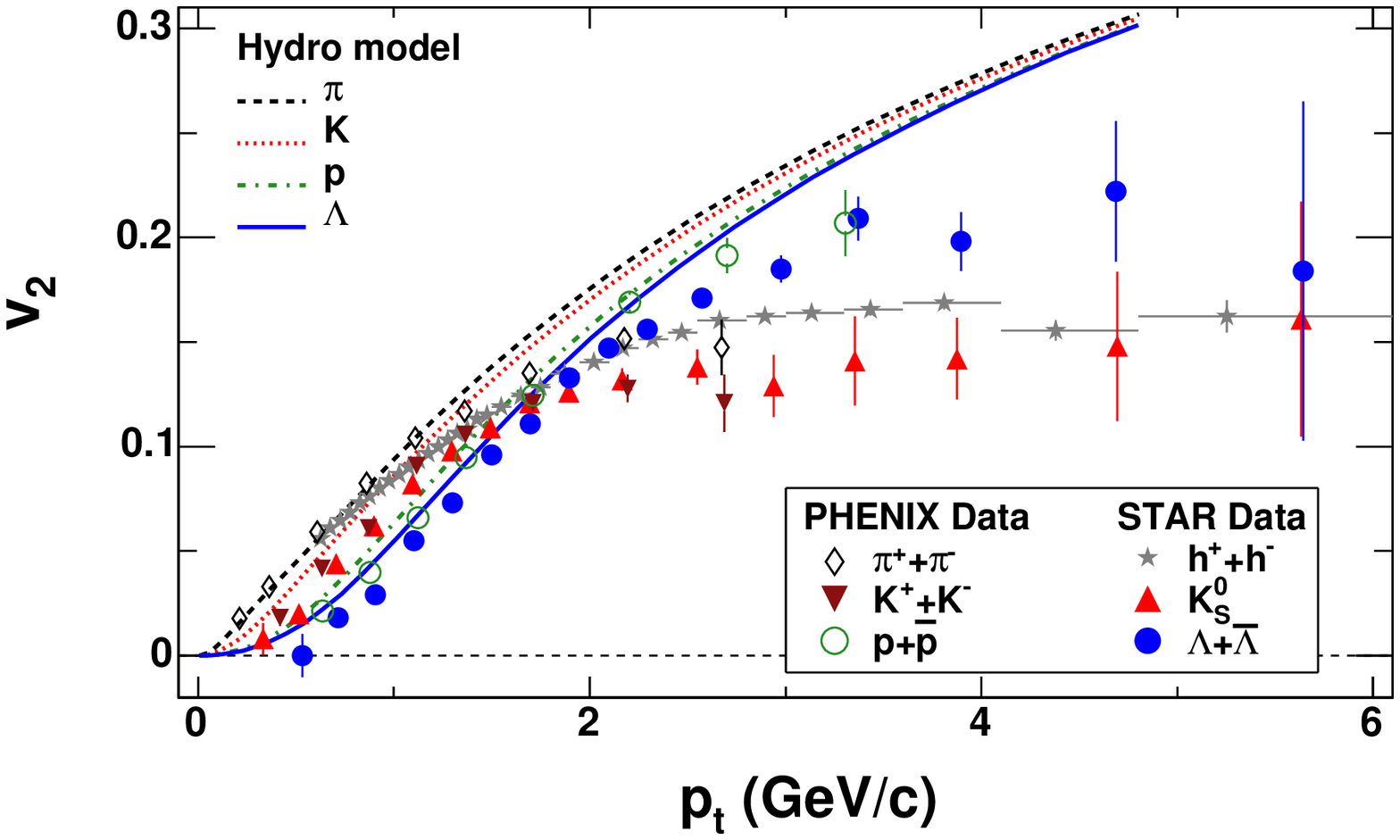,width=6cm}
{\bf FIGURE 2.} \footnotesize{The elliptic flow $\nu_2$ versus $p_T$ for
  different particles together with the hydrodynamics model predictions}
\end{minipage} 

\section {TRANSVERSE MOMENTUM SUPRESSION}
One of the exciting results of the RHIC data is the strong suppression of $p_T$ in central
heavy ion collisions, consistent with the predicted energy loss \cite{8} \cite{9} of the parent light quarks
and gluons when transverse the dense colored medium due to the induced gluon radiation. In
fig. 4 is shown the nuclear attenuation factor 
$R_{AA}(p_T)$.For $\pi^0$ and $\eta$ the data show a suppression factor 5 for $p_T>4$ GeV/c, compared
to the superposition of $NN$ collision, see N. Borghini talk \cite{10}. On the contrary,
$R_{AA}=1$ for direct photons in agreement with perturbative QCD \cite{11} \cite{12}.

\begin{minipage}[t]{6cm}
\centering\epsfig{figure=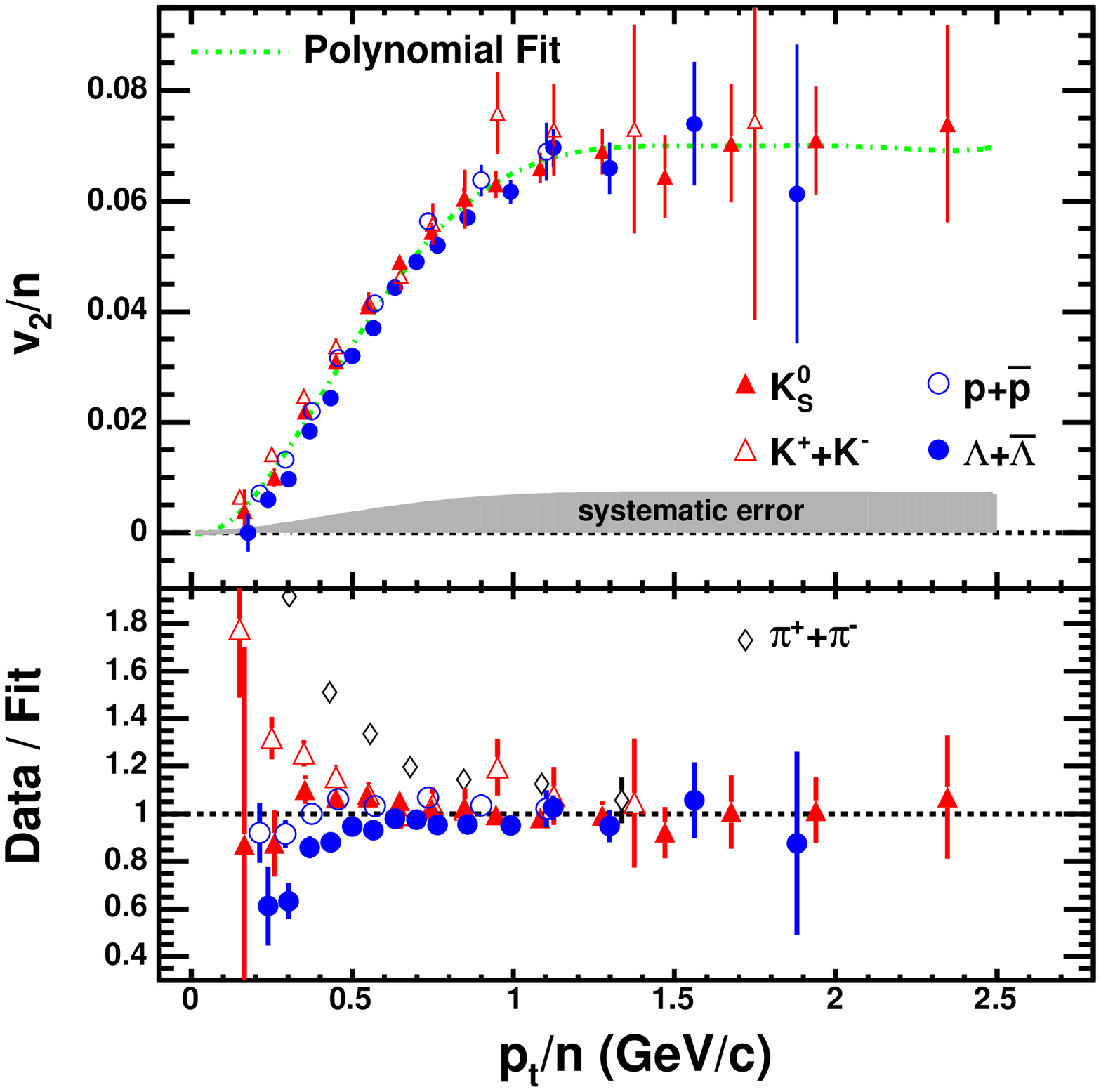,width=6cm}
{\bf FIGURE 3.} {\footnotesize$\frac{\nu_2}{n}$ versus $\frac{p_T}{n}$ being n
the number of quark consituents for different mesons and baryons} 
\end{minipage} 
\hfill
\begin{minipage}[t]{6cm}
\centering\epsfig{figure=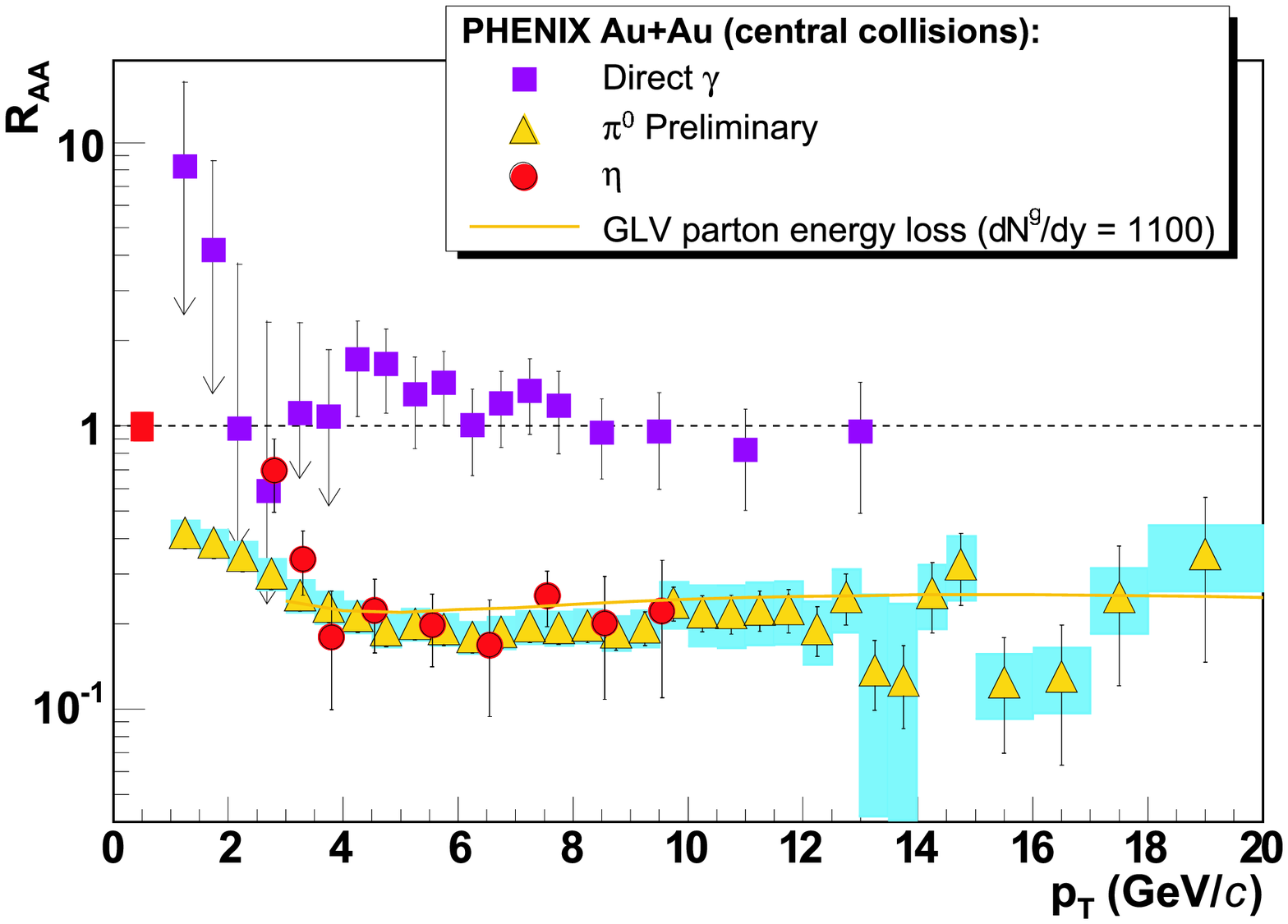,width=6cm}
 {\bf FIGURE 4.} \footnotesize{PHENIX data for central Au-Au collisions of the
 modified nuclear factor $R_{AA}$ as a function of transverse momentum for photons,$\pi⁰$ and $\eta$}
\end{minipage} 

\vspace{1cm}
However, the suppression factor for high $p_T$ electrons from semi-leptonic D
and B decays is as suppressed as the light hadrons in central Au-Au collision \cite{13},see fig. 5, in 
conflict with the prediction of radiative energy loss models. This discrepancy may point
out to a elastic energy loss for heavy quarks. The study of b and c jets at LHC can be
very valuable to clarify this point.

A second exciting phenomena observed at RHIC was the suppression of the back to back jet-like
correlation. Jet-like correlations are measured by selecting the highest $p_T$ trigger hadron
of the event and measuring the azimuthal $\Delta\Phi=\Phi-\Phi_{trig}$ and rapidity
$\Delta\eta=\eta-\eta_{trig}$ distributions of associated hadrons. In $pp$ collisions
a dijet signal appears as two back to back Gaussian peaks at $\Delta\Phi\simeq 0$ (near-side)
and $\Delta\Phi\simeq \pi$ (away-side). On the contrary, the away-side dihadron azimuthal
correlation in central Au-Au collision is clearly suppressed, showing a dip and a double peak
structure \cite{14} at $\Delta\Phi\approx \pi\pm1.1$, for associated hadron in the range
$1\leq p_T \leq 2.5 GeV/c$ (see fig. 6). This double peak structure has been pointed out
as due to the emission of energy from the quenched parton at a finite angle respect to the
jet axis. Such conical configuration can appear if a fast jet moving in a fluid medium
generates a wake of shock wave of Mach Type \cite{15} or Cerenkov Type \cite{16}. In the case
of Mach wave, the characteristic angle $\theta$ of the emitted secondaries determines
the speed of sound, $c_s=cos \theta$. However the double-peak structure of the away-side
correlation is consistent not only with conical emission but also with other scenarios.
In order to distinguish between the different mechanisms, 3-particle azimuthal correlations are
needed. The three particle results reported at this conference \cite{17} as central Au-Au collisions
are consistent with conical emission, but additional studies on the $p_T$ dependence
are needed to emission distinguish between Mach cone shock waves and Cerenkov emission.

\begin{minipage}[t]{6cm}
\centering\epsfig{figure=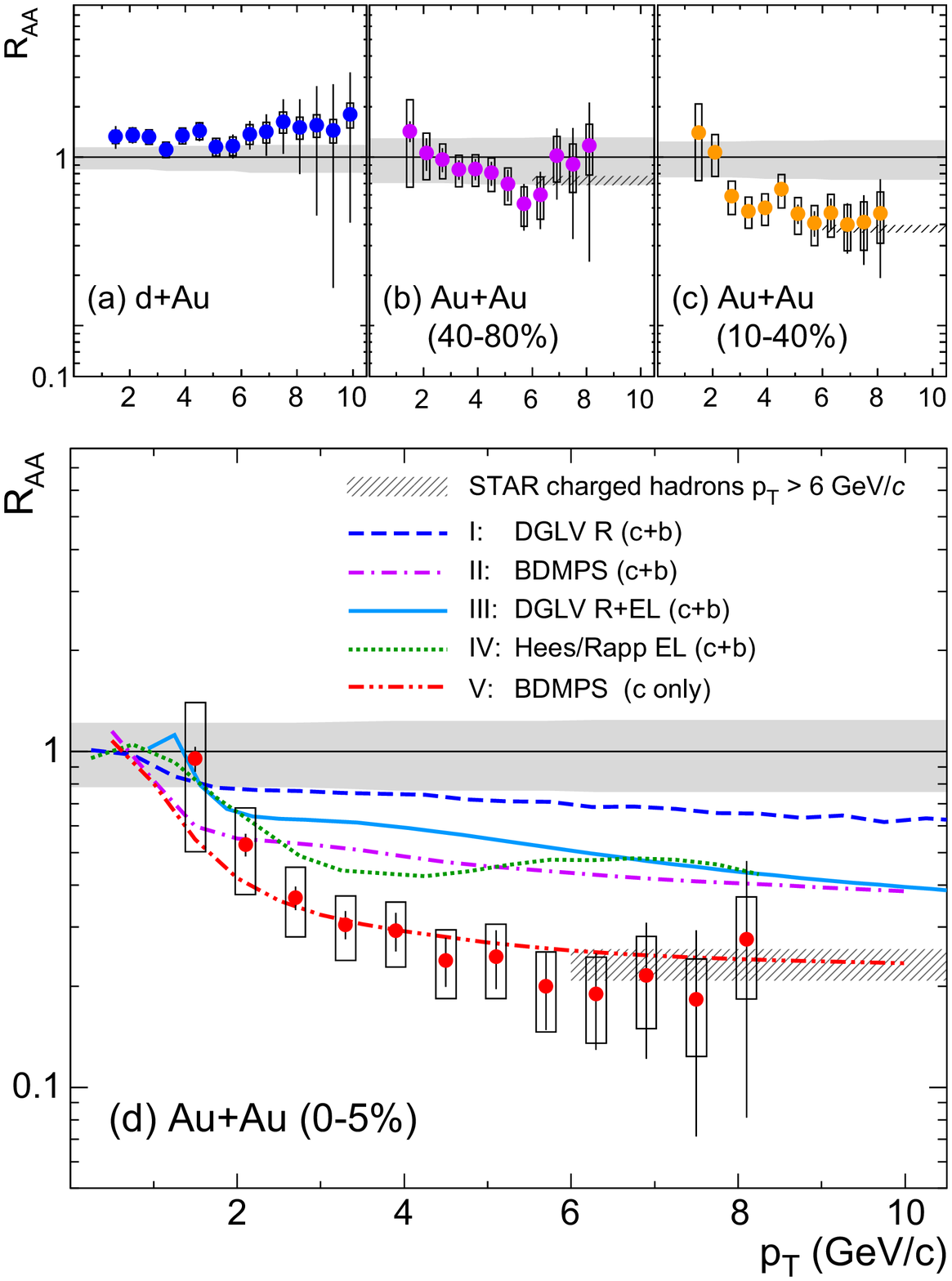,width=4.5cm}
{\bf FIGURE 5.} \footnotesize{$R_{AA}$ versus $p_T$ for nonphotonic electrons for d-Au (a), Au-Au 40-80 (b), Au-Au 10-40 (c)and Au-Au 0-5(d) compared with the STAR data for charged hadrons}
\end{minipage} 
\hfill
\begin{minipage}[t]{6cm}
\centering\epsfig{figure=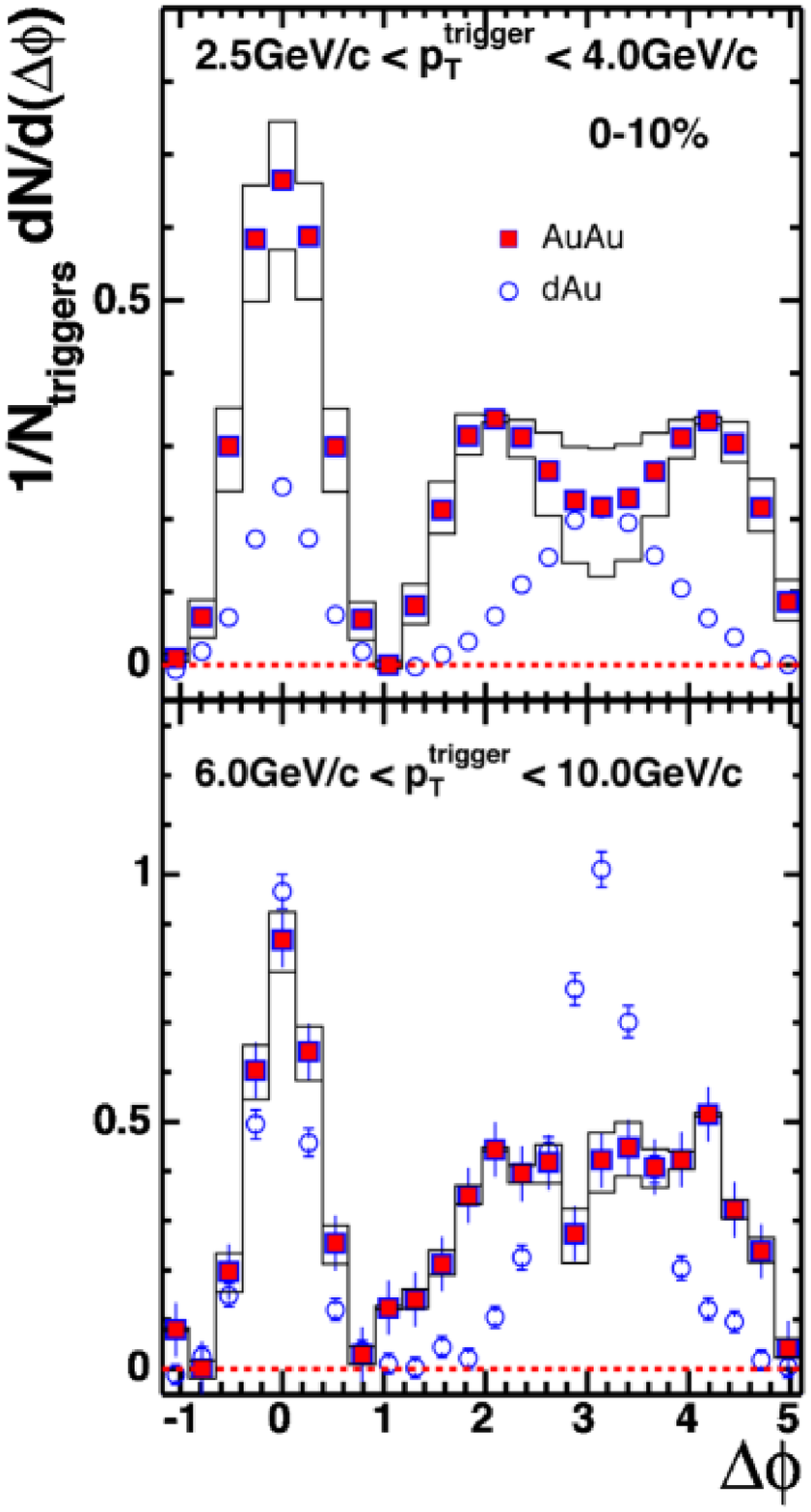,width=4.5cm}
{\bf FIGURE 6.} \footnotesize{STAR data on azimutal distributions of semihard hadrons
  (associated $p_T=1-2.5$ GeV/c) in central Au-Au and d-Au collisions with respect to a trigger hadron measured of 2.5 GeV/c $<p_T<4.0$ GeV/c (top) and 6.0 GeV/c $<p_T<10.0$ GeV/c.}
\end{minipage} 

\section {CHARMONIUM SUPPRESSION}
Early predictions were that the two heavy quarks that would form the bound state would
be screened from each other in the high-density deconfined medium \cite{18}. These states would
melt at different energy densities depending on their size and binding energies. However,
recently, lattice QCD has suggested that the $J/\psi$ would not be screened up to
$T\geq 1.5-2 T_c$. On the contrary other charmonium states would be screened around
$1.1 T_c$ \cite{19}.

The $J/\psi$ suppression at RHIC was predicted to be larger than the observed at SPS by
most of the models. Contrary to this expectation, additional suppression was not
found, fig. 7. In this figure it is also shown the suppression due to normal
absorption using for the absorption cross section the values 1, 3 and 4 mb respectively. The
usual used value of 4.2 mb is higher than the required by $d-Au$ data which is in the
range 1-3 mb. A better understanding of the absorption cross section and in general of
cold nuclear matter effects would be welcome.

We are left with two possible theoretical explanations of the RHIC data, namely regeneration \cite{22}
and sequential dissociation \cite{23} models. In regeneration models a strong dissociation of the
charm pairs due to screening is compensated by the regeneration of bound charm pairs in the
later states of the expansion due to the large production of charmed quarks at high density.
At LHC a large enhancement is predicted.

In the sequential screening model, the $J/\psi$ is not melt at SPS and RHIC energies
as suggested by lattice calculations and the observed suppression comes only from screening of
the higher mass resonances $\psi'$ and $\chi_c$ that though feed down normally provide about
40\% of the $J/\psi$ production. This picture provides a simple explanation for the
similar suppression observed at SPS and RHIC. As at LHC it would be reached temperatures high enough
to dissociate $J/\psi$, it is expected stronger suppression at LHC, contrary to the regeneration model expectation.

On the other hand, the sequential dissociation model relates the observed $J/\psi$ production
to the higher resonance production once the absorption has been substracted. In fact, denoting
by $S_{J/\psi}$ and $S_{\psi '}$, the survival probabilities for the observed total suppression
and for the higher resonances respectively, once the absorption has been substracted

\begin{equation}
        S_{J/\psi}= 0.6+0.4 S_{\psi'}
        \label{ec1}
\end{equation}

Using the absorption cross sections $\sigma_{\psi'}=7.1\pm 1.6 mb$ and 
$\sigma_{J/\psi}=4.3\pm0.3 mb$ we can test the equation (\ref{ec1}) from the data on $J/\psi$ and 
$\psi'$ production, obtaining a good agreement as it is shown in fig. 8.

\begin{minipage}[t]{6cm}
\centering\epsfig{figure=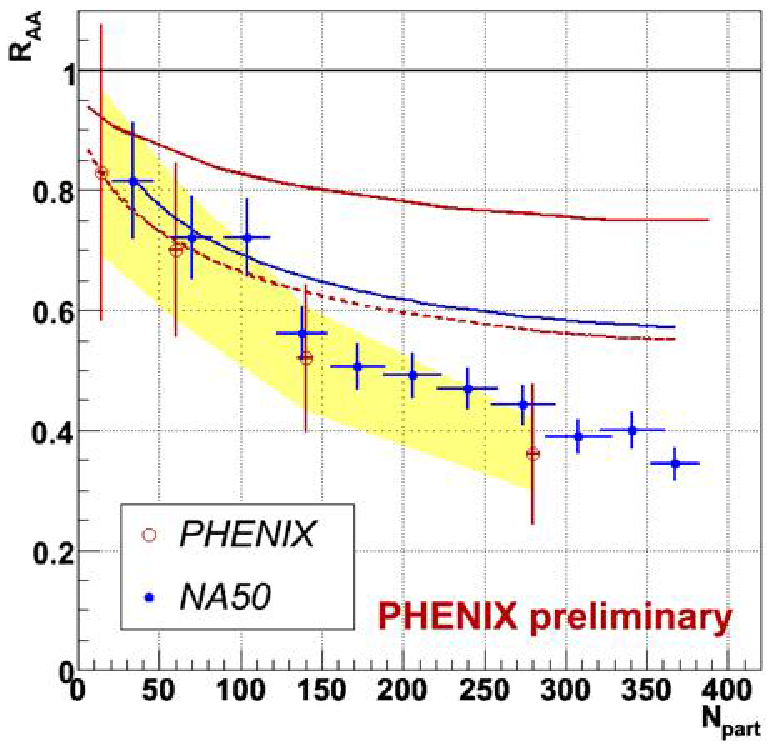,width=6cm}
{\bf FIGURE 7.} \footnotesize{NA50 data together with PHENIX data of the modified nuclear factor $R_{AA}$ for $J/\psi$
  production as a function of the number of participants.}
\end{minipage} 
\hfill
\begin{minipage}[t]{6cm}
\centering\epsfig{figure=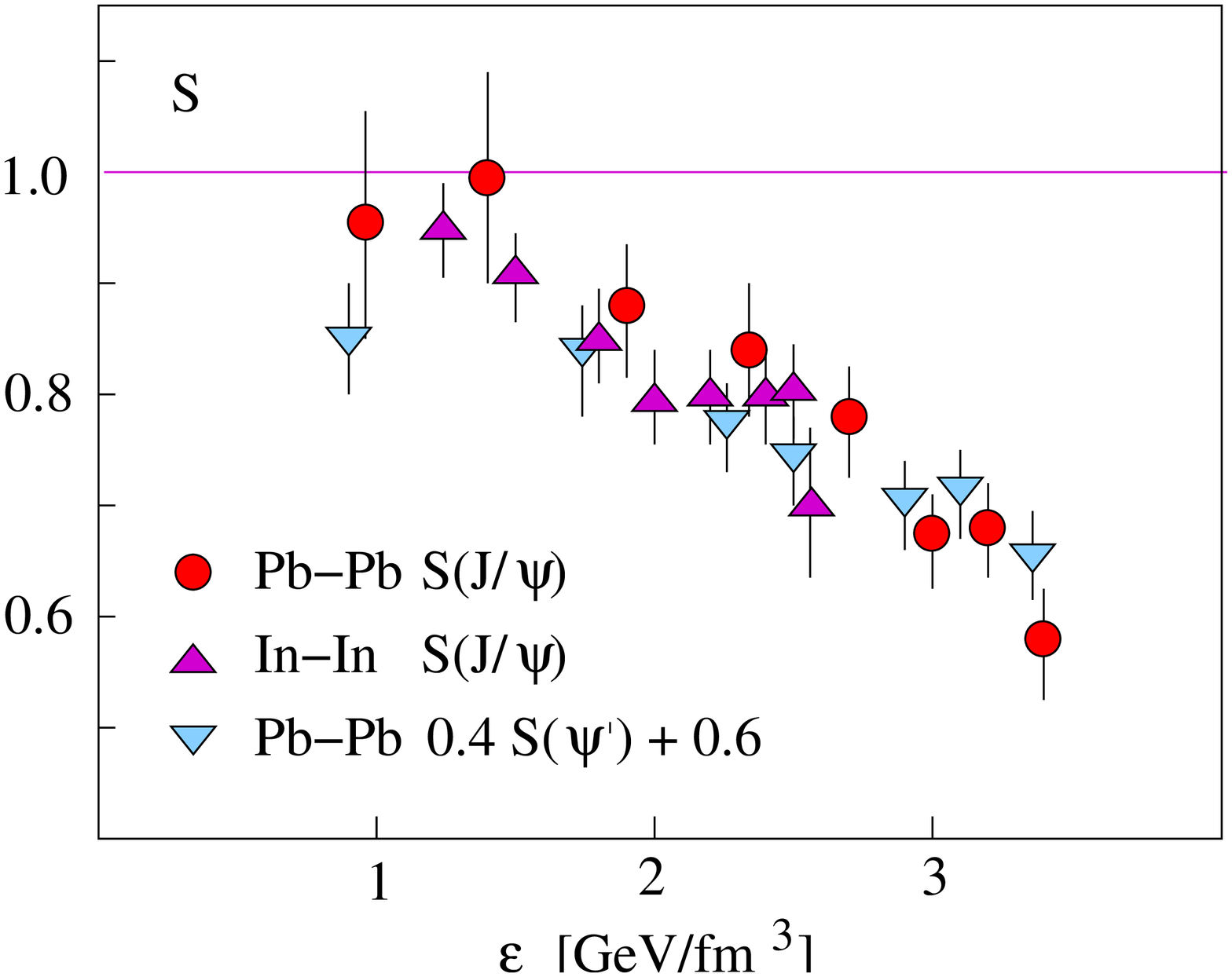,width=6cm}
{\bf FIGURE 8.} \footnotesize{The survival probability of $J/\psi$ for Pb-Pb and In-In
  collisions together with expression (1).}
\end{minipage} 

\section{Phase Transition}

Lattice studies indicates that the deconfinement phase transition at $\mu=0$ is
a cross over \cite{24}, as at $\mu\neq 0$ it is expected to be of first order, in some
point there will be a critical end point. At this conference, were reported studies both
experimental and theoretical on this critical end point. From one side, it was explained
the proposal RHICII \cite{25} which will explore regions of lower energy, looking for
signatures associated to the critical point. In this way, it will join FAIR in the search
of the critical point. On the other hand were reported interesting theoretical results.
B. Kaempfer \cite{26} in a quasi-particle model on of QCD matter study the critical end point
and the equation of state, obtaining with agreement for relevant observables of AA collisions
as the elliptical flow for different particles or the rapidity multiplicity distribution.
Szabo reported the recent results of lattice at $\mu$ and $\mu\neq 0$, in particular the
cross-over at $\mu = 0$. Antonov analytically studied \cite{27} thermodynamics of a heavy quark-antiquark
pair in SU(3) QCD, both below and above the deconfinement critical temperature. He derived
the effective temperature-dependent string tension, which enabled him to calculate internal
energy and entropy of two heavy-light mesons for two flavours. The anomalously large peaks
of these two observables around $T_c$, observed recently on the lattice are well described.

Other interesting aspects of the phase transition (dynamics, flux-tube) were discussed by
G. Krein \cite{28} and G. Kozlov \cite{29}.

\section {Correlations}

In the early stage of heavy ion collision an extended region with large energy density
is produced where quark and gluons degrees of freedom leading to a new partonic phase 
of matter. In the subsequent evolution, the system dilutes and cools down, hadronizes and
finally decays into observed hadrons. These hadrons carry only indirect information about the
early stage of the collision. Results as the elliptic flow discussed before suggest that a
deconfined phase starts in the early stage of the reaction. The study of correlation
and fluctuations can provide additional information on the reaction mechanism. For these
reasons correlations between oppositely charge particles and multiplicity and transverse
momentum fluctuations have been measured in the last years.

The balance function \cite{30} measured the correlation of the oppositively charged particles.
The width of the balance function $<\Delta y>$
is sensitive to the hadronization time. 

If the system produced in a heavy-ion collision has undergone a partonic phase, the hadronization
will occur at later time and therefore the temperature will be lower and the diffusive 
interaction with other particles will be lesser than those in the direct hadronization \cite{31}.
A delayed hadronization implies stronger correlation in rapidity for the charged particles
and therefore a narrower balance function. Indeed a narrowing of the balance function is
observed with increasing size of the colliding nuclei by the NA49 and STAR Collaboration \cite{32}.
The Hijing model as well as shuffled events retaining only correlations
from global charge conservation do not show any decrease of the width. However, other
models without delayed hadronization can describe the data \cite{33}. Notice, that the integral
of the balance function is related to the event by event charge fluctuations, which are
expected to be suppressed in a QGP \cite{34}.

The multiplicity correlations have been studied by the NA49 Correlations \cite{35}. In fig. 9,
the scaled variance of negative particles is shown as a function of the number
of projectile participants together with the results of three different string models.
It has been pointed out that for peripheral collisions a significant contribution comes
from fluctuations in the number of target participants at fixed projectile participant number,
what means that the projectile and target hemisphere are connected. The strings models of
fig. 9 are based on Fritiof model where the strings are stretched between partons of the
same excited hadron and therefore there is not connection between hemispheres. This does
not happen in string models like, dual parton model, quark gluon string model and Venus where there
is color exchange and the strings are stretched between partons of the projectile and
the target. 

\begin{minipage}[t]{6cm}
\centering\epsfig{figure=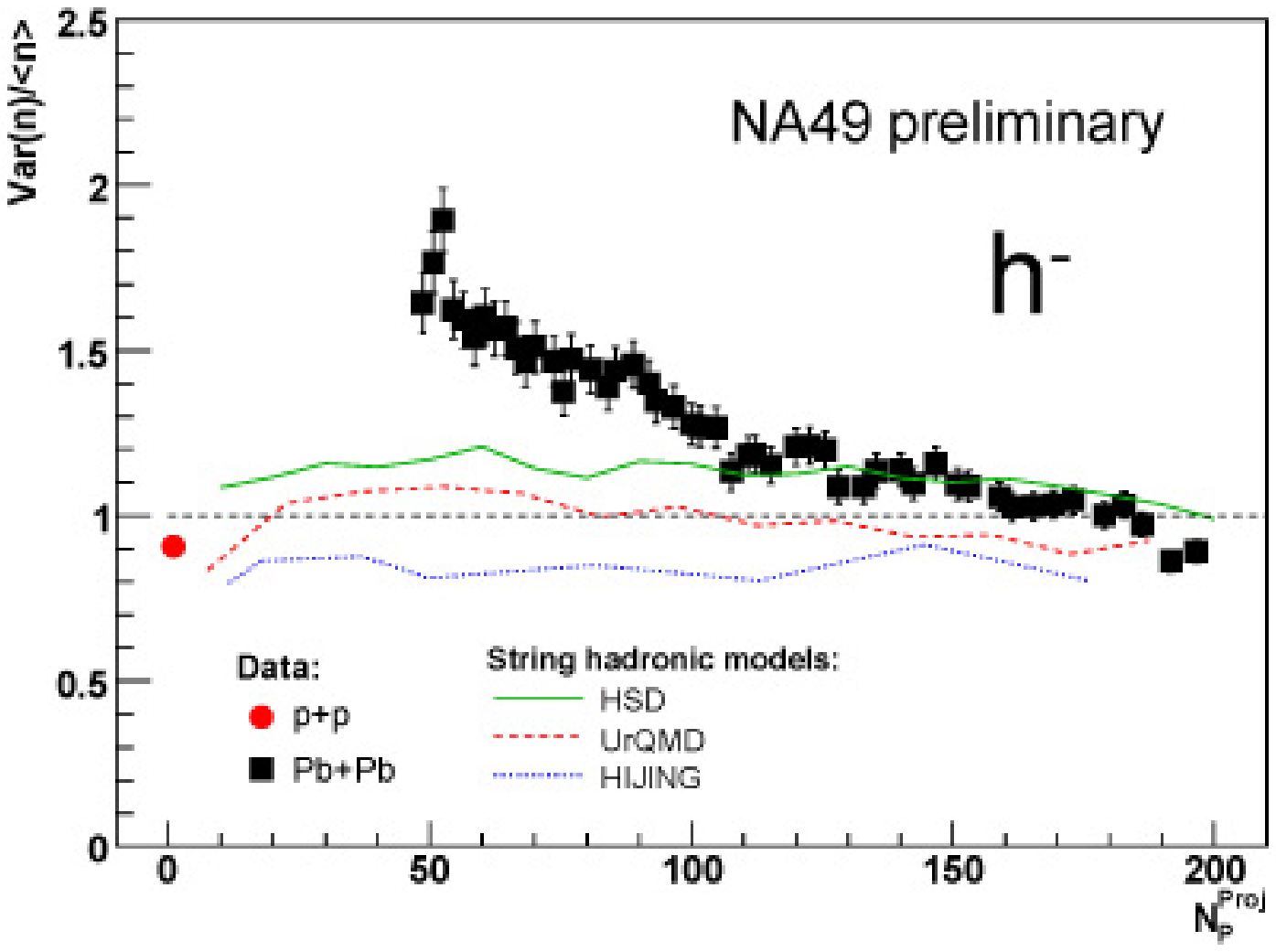,width=6cm}
{\bf FIGURE 9.}\footnotesize{NA49 experimental data on the scaled variance of negative
  multiplicity of Pb-Pb collisions as a function of the number of projectile
  participants together the predictions of different non color exchange string models.}
\end{minipage} 
\hfill
\begin{minipage}[t]{6cm}
\centering\epsfig{figure=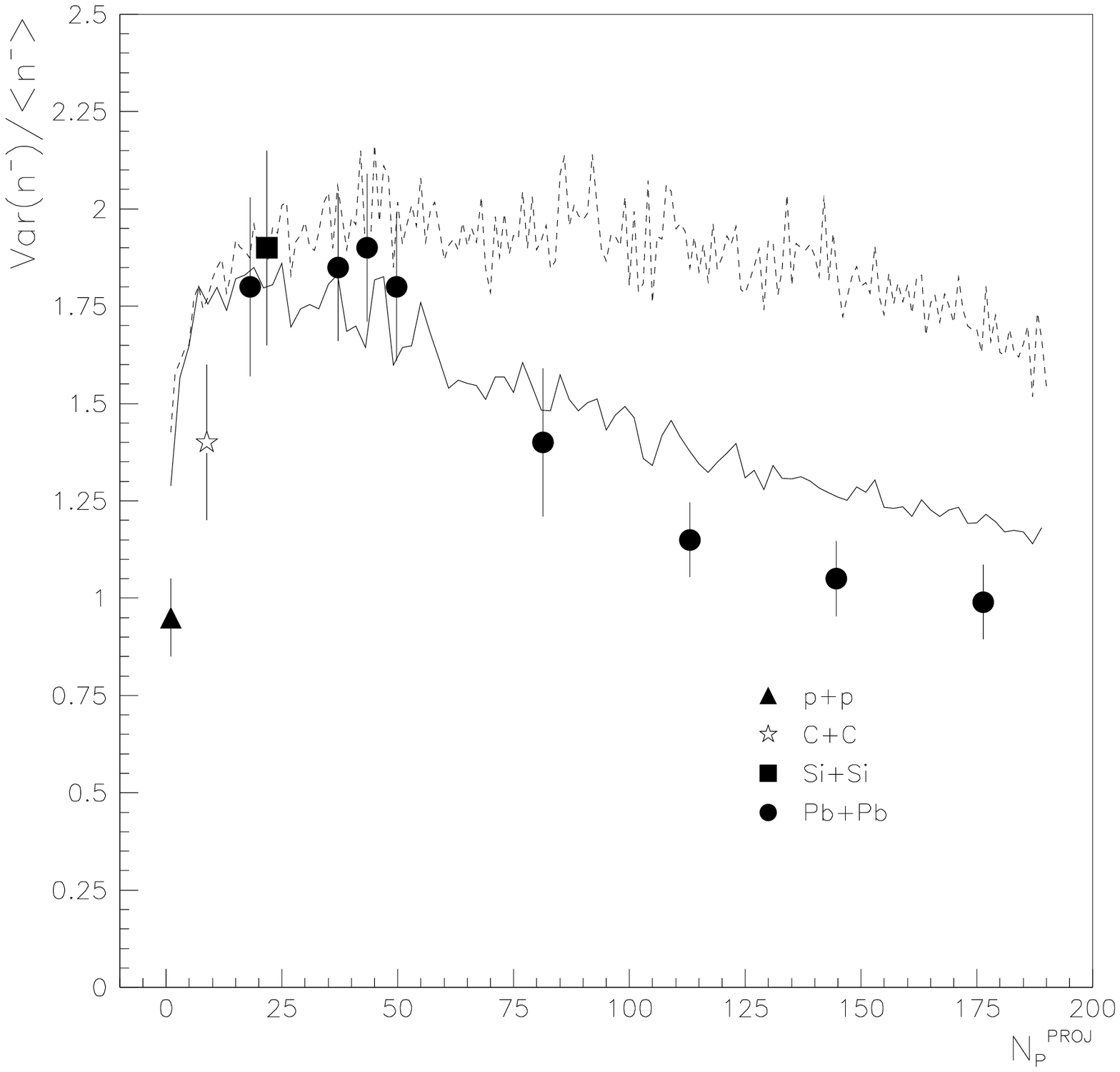,width=6cm}
{\bf FIGURE 10.} \footnotesize{NA49 preliminary data on scaled variance of negative
  multiplicity as a function of the number of projectile participants compared with the result of percolation of strings.}
\end{minipage}

\vspace{1cm}
In fig. 10 it is shown the result of a percolation color sources model \cite{37} together
with previous NA49 data. It is seen that contrary to the above string models in this case
reproduced the general trend of data. The percolation framework is able to describe
the dependence on the centrality of the transverse momentum fluctuations \cite{38}. In this approach
the strings stretched between partons of the projectile and target. In the transverse space these
color strings are seen as small circles, with $r_0\simeq 0.2-0.3 fm$. With growing energy
and/or atomic number, the number of strings grows, starting to overlap forming clusters.
Each cluster decay into particles with a mean multiplicity and mean transverse momentum which
depends on the number of strings of the cluster and the total area of the cluster.
These dependence are essentially determined by the Schwinger mechanism and the color field of
each cluster. At a certain critical density, a macroscopic cluster appear which marks the
percolation phase transition \cite{39}. The fluctuations are understood as follows: At low density,
most of the particles are produced by individual strings with the same mean multiplicity
and mean $p_T$, therefore small fluctuations.
At large density above the critical point essentially there will be only are cluster
and therefore the fluctuations are small. The maximum of fluctuations corresponds to the
largest number of clusters with different size and number of strings.

The percolation of strings predicts that the long range rapidity correlation measured by
$D^2_{FB}=<n_B n_F>-<n_F><n_B>$ (between forward F and backward B the rapidity gap should
be larger than 1-1.5 to eliminate short range correlations) increases with centrality,
being much less than what is expected from superposition models \cite{40}. This is in
agreement with STAR preliminary data \cite{41}.

In Color Glass Condensate(CGC)\cite{42} we expect also that $D^2_{FB}$ grows with the
centrality \cite{43}. In fact, the main contribution is given by

\begin{equation}
        <\frac{dN}{dy_1}\frac{dN}{dy_2}>\simeq 
        <\left(\frac{dN}{dy}\right)^2>=
        \sim \frac{1}{\alpha_s^2} \pi R^2 Q^2_S
        \label{ec2}
\end{equation}

As centrality increases, $\alpha_s$ decreases and (\ref{ec2}) grows.

\section{Thermalization}

It is clear that a fast thermalization of the partonic is required. 
We learned at beginning of the conference \cite{44} that this can be achieved naturally in
the CGC approach. In this approach, the collision of two heavy ion, which are gluon saturated in
the initial state, develop strong longitudinal chromoelectric fields \cite{45}, which via the
Schwinger mechanism produce particles with a thermal spectrum due to the fluctuations
of the color field. The temperature T of this spectrum is related to the saturation
momentum $Q_s, T\simeq Q_s/2\pi$. The thermalization time is $\tau\sim 1/Q_s$.

As it has pointed out \cite{44}, a similar picture arises in percolation \cite{46}. In this case, the
critical density $\eta_c$ for the non-thermal percolation phase can be related to the critical
temperature, $T_c=\frac{<p_T>_1}{\sqrt{2 F(\eta_c)}}$, where $<p_T>_1$ is
the mean transverse momentum of particles produced in one single string (essentially the
string tension) and $F(\eta_c)=\sqrt{\frac{1-e^{\eta_c}}{\eta_c}}$ has geometrical
origin and has to do with the fraction of total avalaible area occupied by the cluster
$(1-e^{-\eta_c})$. The shear viscosity is $<p_T>_1 F(\eta)L$ is also determined by the
same factor . L is the longitudinal extension, $L \simeq 1 fm$.
For reasonable values of $<p_T>_1 \simeq 200 MeV$ and $\eta_c \simeq 1.2-1.5$, it is
obtained $T_c=170-180 MeV$ and a very low shear viscosity.

In conclusion, theoretical and experimental progress have been achieved in the understanding
of deconfinement as it has been reported in the different talks of this session. The future
experiments of LHC, FAIR and we expect that also RHIC II will let us to go on this 
understanding, and clarify some of the questions of the field.

%\subsection{<A subsection>}

%\subsubsection{<A subsubsection>}

\begin{theacknowledgments}
  
We thank the organizers for such a nice meeting and Y. Foka, J. Rafelski and M. Leith
for helping in the configuration of this talk. The work has done under contract
FPA2005-01963 of SPAIN and the support of Xunta de Galicia.

\end{theacknowledgments}

\end{document}